\documentclass{llncs}
\usepackage{graphicx}
\usepackage{algorithm}
\usepackage{algorithmic}
\usepackage{subfigure}
\usepackage{amsmath} 

\begin{document}
\title{GraCT: A Grammar based Compressed representation of Trajectories \thanks{\scriptsize{This work was funded in part by European Union’s Horizon 2020  Marie Sk{\l}odowska-Curie grant agreement No 690941; Ministerio de Econom\'{\i}a y Competitividad under grants [TIN2013-46238-C4-3-R], [CDTI IDI-20141259], [CDTI ITC-20151247], and [CDTI ITC-20151305]; Xunta de Galicia (co-founded with FEDER) under grant [GRC2013/053]; and Fondecyt Grant 1-140796, Chile. }	}}

\author{Nieves R. Brisaboa\inst{1} \and Adri\'an G\'omez-Brand\'on\inst{1} \and Gonzalo Navarro\inst{2} \and Jos\'e R. Param\'a\inst{1}
}
\institute{Depto. de Computaci\'on,
 Universidade da Coru\~na, Spain \\
\email{\{brisaboa, adrian.gbrandon, jose.parama\}@udc.es}
\and
Dept. of Computer Science, University of Chile, Chile. 
\email{gnavarro@dcc.uchile.cl} }

\maketitle

\begin{abstract}

%

We present a compressed data structure to store free trajectories of moving objects (ships over the sea, for example) allowing spatio-temporal queries. Our method,  GraCT, uses a $k^2$-tree to store the absolute positions of all objects at regular time intervals (snapshots), whereas the positions between snapshots are represented as logs of relative movements compressed with  Re-Pair. Our experimental evaluation shows important savings in space and time with respect to a fair baseline.

\end{abstract}

\section{Introduction}

After more than two decades of research on moving objects, this field still presents interesting problems that represent a topic of active research. The renewed interest to represent and exploit data about moving objects is mainly due to the new context in which large amounts of
data (from, for example, cellular phones informing about the GPS coordinates of their  position in real time) need to be
stored and analyzed. Therefore, new big data sets and new application domains demand more efficient technology to manage moving objects.

Traditional spatio-temporal  indexes can be classified into two families, {\em space-based} indexes and {\em trajectory-based}
indexes. Each type of index is adapted to answer different types of queries. Indexes in the first family usually are modifications of the classical spatial R-tree, like for example the RT-tree \cite{Xu90}, the HR-tree \cite{NascimentoS98}, the 3DR-tree \cite{Vazirgiannis1998},  the  MV3R-Tree \cite{PapadiasT01}, or the SEST-Index \cite{GutierrezNRGO05}. Those indexes efficiently answer queries which return the ids or the number of objects into a given spatial region at a specific time instant (time-slice queries) or at a specific time interval (time-interval queries), but they cannot efficiently return the position of an object at a time instant or which was its trajectory\footnote{We informally define trajectory as a list of positions in consecutive time instants.} during a time interval.

The second family of indexes were designed to improve the management of trajectories, like   SETI \cite{ChakkaEP03}, the CSE-tree
\cite{WangZXM08}, and trajectory splitting strategies \cite{hadjieleftheriou2002efficient,rasetic2005trajectory}. Those indexes can describe trajectories of individual objects but cannot answer efficiently time-slice or time-interval queries over objects in a specific region of the space.

Those indexes maintain the bulk of the data on disk, while the index structures
reside
in main memory. They rarely use compression to reduce disk or memory usage, or
to reduce the disk transfer time.
%










%











In this paper we introduce an in-memory representation called {\em Grammar based Compressed representation of Trajectories (GraCT)}. GraCT is a trajectory-oriented technique, that is, it belongs to the second family. However, it structures the index into snapshots of the objects taken at regular time instants, and logs of their movements between snapshots. This allows GraCT to efficiently answer \textit{time-slice} and \textit{time-interval} queries as well, by processing the logs between two snapshots. Besides, GraCT represents data and index together, and uses grammar compression on the logs. This not only reduce the size of the representation, but also the nonterminals are enriched to allow processing long parts of the log files without decompressing them, and {\em faster} than with a plain representation. Its space savings allow GraCT fitting much larger datasets in main memory, where they can be queried much faster than on disk.

\section{Background}

\subsection{\large{$\boldsymbol{K^2}$}-\bf{tree}}\label{k2}

The $k^2$-tree is a compact data structure originally designed for
representing Web graphs in little space, allowing its manipulation directly in compressed form  \cite{ktree}.
The $k^2$-tree is used to represent the adjacency matrix of the graphs, and it can also be used to represent any type of binary matrices.

\begin{figure}[t]
\begin{center}

\includegraphics[scale=0.15]{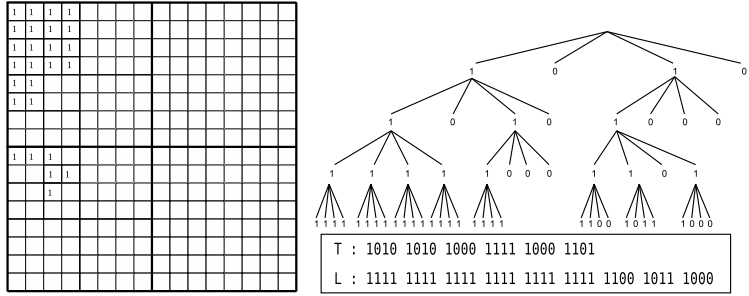} 

\vspace*{-5mm}
\end{center}
\caption{Example of binary matrix(left) and resulting $k^2$-tree conceptual representation (right up), and the compact representation (right down), with $k=2$.}
\label{fig:example}
\end{figure}

The $k^2$-tree is conceptually a non-balanced $k^2$-ary tree built from a binary matrix by recursively subdividing the  matrix into
$k^2$ submatrices of the same size. It starts by subdividing the original matrix into
$k^2$ submatrices of size $n^2/k^2$, being $n \times n$ the size of the matrix. The submatrices are ordered from left to right and from top to bottom. Each of those submatrices  generates a child of the root node whose value is $1$ if there is at least one $1$ in the cells of that submatrix, and $0$ otherwise. The subdivision proceeds recursively for each child with value $1$ until it reaches a submatrix full of 0s, or it reaches the cells of the original
matrix (i.e., submatrices of size $1\times 1$). Figure~\ref{fig:example} shows an example of this subdivision (left) and the resulting conceptual $k^2$-ary tree (right up) for $k=2$.

Instead of using a pointer-based representation, the $k^2$-tree is compactly stored using two bitmaps $T$ and $L$ (see Figure \ref{fig:example}).
\textit{T}  stores all the bits of the $k^2$-tree except those in the last level. The bits are placed following a levelwise traversal: first the $k^2$ binary values of the root node, then the values of the second level, and so on. $L$ stores the last level of the tree. Thus, it represents the value of original cells of the binary matrix.

It is possible to obtain any cell, row, column, or region of the matrix very efficiently, by just running $rank$  and $select$ operations \cite{Jacobson89} over the bitmap $T$: $rank_b(T,p)$ is the number of occurrences of bit $b \in \{0,1\}$ in $T$ up to position $p$, and $select_b(T,j)$ is the position in $T$ of the $j$th occurrence of the bit $b$. For example, given a value 1 at position $p$ in $T$, its $k^2$ children will start at position $p_{children}=rank_1(T, p) \times k^2$ of $T$, except when the position of the children of a node returns a position $p_{children} > |T|$; in that case we access instead $L[p_{children}-|T|]$ to retrieve the actual value of the cells. Similarly, the parent of a position
$p$ in $T:L$ is $q-(q \!\!\mod k^2)$, where $q = select_1(T,\lfloor p/k^2\rfloor)$, and 
$q \!\!\mod k^2$ indicates which is the submatrix of $p$ within its parent's.

\subsection{Re-Pair}

\begin{figure}[t]
\begin{center}
\includegraphics[scale=0.33]{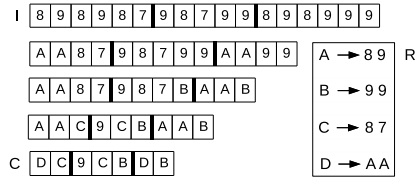}
\vspace*{-2mm}
\caption{An example of Re-pair compression.}\label{re-pair}
\end{center}
\end{figure}

Re-pair~\cite{larsson2000off} is a grammar-based compression method. Given a
sequence of integers $I$ (called {\em terminals}), it proceeds as follows: (1) it obtains  the most frequent pair of integers $ab$ in $I$, (2) it adds the rule $s \rightarrow ab$ to a dictionary $R$, where $s$ is a new symbol not appearing in $I$ (called a {\em nonterminal}), (3) every occurrence of $ab$ in $I$ is replaced by $s$, and (4) it repeats steps 1-3 until every pair in $I$ appears only once (see Figure \ref{re-pair}). The resulting sequence after compressing $I$ is called $C$. Every symbol in $C$ represents a phrase (a sequence of 1 or more of the integers in $I$). If the length of the represented phrase is 1, then the phrases consists of an original (terminal) symbol, otherwise it is a new (nonterminal) symbol. Re-Pair  can be implemented in linear time, and a phrase can be recursively expanded in optimal time (that is, proportional to its length).

\section{Our approach}
 
GraCT represents moving objects that follow free trajectories on the space. 
We consider the time as discrete, therefore each time instant actually corresponds to a short period of time. We assume that in each time instant, each object informs its position (e.g., international regulations require that ships inform their GPS position at regular intervals). 
We use a raster model to represent the space, therefore the space is divided into cells of a fixed size, and objects are assumed to fit in one cell. The size of the cells and the period used to sample the time are parameters that can be adapted to the specific domain.

 Every  $s$  time instants, GraCT uses a data structure based on $k^2$-trees to represent the absolute positions of all objects. We call those time instants snapshots. The distance  $s$ between snapshots is another parameter of the system. Between two consecutive snapshots the trajectory of each moving object is represented as a log, which is an array of movements, that is, relative positions with respect to the previous time instant. 

\vspace{-0.4cm}
\subsubsection{Snapshots}

\begin{figure}[t]
\begin{center}

\includegraphics[scale=0.22]{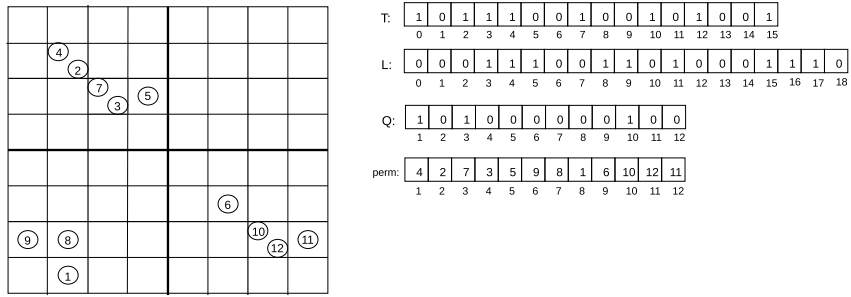}
\end{center}
\vspace*{-5mm}
\caption{The position of objects in the space (left), and the representing snapshot (right).}\label{snapshot}
\end{figure}

Each snapshot uses a $k^2$-tree where a cell set to 1 indicates that one or more objects are placed in that cell, whereas a 0 means that no object is in that cell.
However, we still need to know which objects are in a cell set to 1.
Observe that each 1 in the binary matrix corresponds to a bit set to 1 in the bitmap $L$ of the $k^2$-tree. We store the list of object identifiers corresponding to each of those bits set to 1 in an array, where the objects identifiers are sorted following the order of appearance in $L$.
We call that array {\em perm}, since that array is a permutation \cite{Knuth91a}. In addition,
we need a bitmap, called \textit{Q}, aligned with \textit{perm}, that informs with a 0 that the object identifier aligned in \textit{perm} is the last object of a  leaf, whereas a 1 signals that more objects exist.
Observe in Figure \ref{snapshot}, the object identifiers corresponding to the first 1 in $L$ (which is at position 3 of $L$) are stored starting at position 1 of $perm$. In order to know how many objects are in the corresponding cell, we access $Q$ starting at position 1 searching for the first 0, which is at position 2, therefore there are two objects in the inspected cell. By accessing positions 1 and 2 of {\em perm}, we obtain the object identifiers 4 and 2. Now, in position 3 of {\em perm} starts the object identifiers corresponding to the second 1 in $L$, and so on.



With these structures used to represent the absolute positions of all
the moving objects at snapshots we can answer two types of queries:

\begin{itemize}
\item \textit{Find the objects in a given cell}: First,
using the procedure shown in Section \ref{k2}  to navigate downwards
the $k^2$-tree, we traverse the tree from the root  until reaching
the position $n$ in $L$ corresponding to that cell. Next,  we count
the number of 1s in the array of leaves $L$ until the position $n$;
this gives us the number of leaves with objects up to the $n^{th}$
leaf, $x = rank_1(L, n)$. Then we calculate the position of the
$(x-1)$th 0 in $Q$, which indicates the last bit of the previous leaf
(with objects), and we add 1 to get the first position of our leaf,
$p = select_0(Q, x-1)+1$. Then $p$ is the position in $perm$ of the first
object identifier corresponding to the searched position. From $p$,
we read all the object identifiers aligned with 1s in $Q$, until we
reach a 0, which signals the last object identifier of that leaf.

\item \textit{Find the position in the space of a given object}.  First,
we need to obtain the position $k$ in \textit{perm}  of the searched
object. In order to avoid a sequential search over \textit{perm} to
obtain that position, we add additional structures to compute cells of
the inverse permutation of {\em perm} \cite{munro2012succinct}.
Then, we have to find the leaf in $L$ corresponding to the
$k^{th}$ position of $perm$. For this sake, we calculate the number
of leaves before the object in position $k$ of {\em perm}, that is,
we need to count the number of 0s until the position before $k$, $y
= rank_0(Q, k-1)$. Then we find in $L$ the position of the
$(y+1)^{th}$ 1, that is, $select_1(L, y+1)$. With that position of
$L$, we can traverse the $k^2$-tree upwards in
order to obtain the position in the space of that cell, and thus
the position of the object.
\end{itemize}

\vspace{-0.35cm}
\subsubsection{Log of relative movements}

The changes that occur between snapshots are tracked using a log file
per object.
The use of snapshots
and logs is not new \cite{Worboys05}, but in previous works log
values are  stored according the appearance of ``events'' (such as
objects that appear in or disappear from an area).

\begin{figure}[t]
\begin{center}

\includegraphics[scale=0.4]{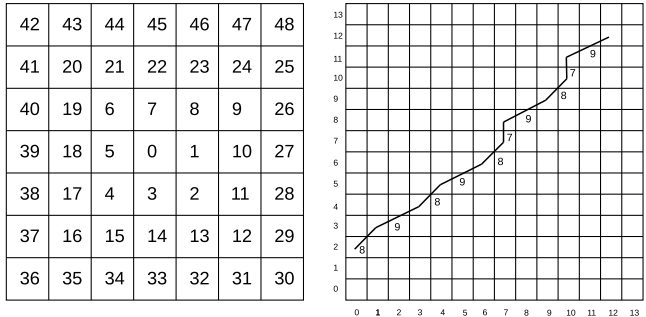}
\end{center}
\vspace*{-5mm}
\caption{Encoding object's movements.}\label{espiral}
\end{figure}

The log stores relative movements with respect to the last
known position of an object, that is, to its position in the preceding
time instant. Objects can change their positions along the two
Cartesian axes, so every movement in the log can be described with two
integers. Instead, in order to save space, we encode the two
values with a unique positive integer. For this sake, we enumerate
the cells around the actual position of an object, following a
spiral where the origin is the initial object position, as it is
shown in Figure \ref{espiral} (left). Let us suppose that an object
moves with respect to the previous known position one cell to the East
in the x-axis, and one cell to the North in the y-axis. Instead of
encoding the movement as the pair (1,1), we encode it as an 8. In
Figure \ref{espiral} (right) we show the trajectory of an object
starting at cell (0,2). Each number indicates a movement between two
consecutive time instants. Since most relative movements involve short
distances, this technique produces a sequence of usually small numbers.

Sometimes real objects stop emitting signals during periods of time.
This forces us to add two new possible movements inside a log:
\textit{relative reappearance} and \textit{absolute reappearance}.
We reserve two codewords to signal these events. We use a relative
reappearance when an object disappears and reappears between the
same snapshots, and an absolute reappearance otherwise. Relative
reappearances are followed by the time elapsed from the disappearance
and a relative movement from that time instant, whereas absolute
reappearances are followed by the  number of time instants that
elapsed since the disappearance and the absolute values of the $(x,y)$
coordinates of the new position of the object. 

\section{Compressing the log}

The log not only saves much space compared to using $k^2$-trees for
every instant, but it also offers important opportunities for further
compression. A first choice is statistical compression, since as said,
most movements are short-distanced and thus our spiral encoding uses
mostly small numbers. We exploit this fact using $(s,c)$-Dense Codes
(SCDC) \cite{RodrguezBrisaboa07}, a very fast-to-decode statistical 
compressor that has a low redundancy over the zero-order empirical
entropy of the sequence. We will use this approach as fair baseline. 

The second approach, which gives the title to this paper, uses grammar
compression on the set of all the log files. Our aim is to exploit the 
fact that there are typical trajectories followed by many objects, which
translate into long sequences of identical movements that grammar
compression can convert into single nonterminals. This includes, in
particular, long straight trajectories in any direction.

\vspace{-0.35cm}
\subsubsection{ScdcCT: Using SCDC for compressing the logs}
The size of the cells and the time elapsed between consecutive time
instants must be carefully chosen to represent properly the typical
speed  of moving objects, so that short movements to contiguous
cells are more frequent than movements to distant cells. Instead of
sorting the spiral codes by frequency, we will simply assume that
smaller numbers are more frequent than larger ones. Since the $(s,c)$-codes
depend only on the relative frequency of the symbols, we do not need to
store any statistical model. Still, we will use the frequencies to optimize
$s$ and $c$ in order to minimize the space usage.

\vspace{-0.35cm}
\subsubsection{GraCT: Using Re-Pair for compressing the logs}

Moving objects spend most of the time either stopped or moving
following a specific  course and speed. In both cases, the logs will
present longs sections with numbers representing the same or
contiguous values of the spiral. For example,  the moving object in
Figure \ref{espiral} follows a NE trajectory moving one or
two cells per time instant. Therefore its log represents the series
of relative movements 8,9,8,9,8,7,9,8,7,9; see the array $I$
of Figure \ref{re-pair}. Those series of similar
movements are very efficiently compressed using a grammar compressor such
as Re-Pair. To avoid having to decompress the log before processing it,
we enrich the rules of the grammar $R$ with further data apart from the
two symbols to be replaced.
Specifically, each rule in $R$ will have the following information:
$s \rightarrow a, b, \#t, x, y, MBR$, where $s$, $a$ and $b$ are the
components of a normal rule of Re-Pair, $\#t$ is the number of
instants covered by the rule, $(x,y)$ are the  relative coordinates
of the final position of the object after the application of the
rule, and $MBR$ is the minimum bounding rectangle enclosing the
movements of the rule.

For example, the rules of Figure \ref{re-pair} are enriched as follows. The
first rule of $R$ is $A \rightarrow 8,~9, ~2,~ (3,2), ~(0,0,3,2)$:
$8$ and $9$ are the substituted symbols, $2$ indicates that the rule
represents a sequence of $2$ movements, $(3,2)$ indicates
the position of the object after the application of the rule if we
start at $(0,0)$, and the last four values are two points
defining a rectangle that encloses all the movements encoded by the
rule.  The other rules are  $B \rightarrow 9,~9, ~2, ~(4,2),
~(0,0,4,2)$, $C \rightarrow 8,~7, ~2,~ (1,2), ~(0,0,1,2)$, and $D
\rightarrow A,~A, ~4,~ (6,4), ~(0,0,6,4)$.

Thanks to this additional information, to obtain the position of an
object at any time instant between two snapshots, the nonterminal
symbols of array $C$ do not need to be decompressed in most cases.
Assume we want to know the position of the object at
the 5$^{th}$ time instant, which is when the object in Figure
\ref{espiral} (right)  is at position (7,7). The preceding snapshot
informs that the absolute position of the object at the beginning of
the log is (0,2). Next, we inspect the log (the $C$
array of Figure \ref{re-pair}) from the beginning. The first value
is a $D$. The enriched rule indicates that 
such symbol represents 4 time instants, and after it, the object is
displaced 6 columns to the East and 4 rows to the North, that is,
starting at (0,2), after the application of this rule, the object
will be at (6,6). Since our target time instant is later than the
final time instant of this rule, we do not have to decompress it,
and this is the usual case. The next symbol is $C$, which lasts 2
time instants. This would take us to time instant 6, but this
surpasses our target time instant(5). Therefore, in this case, that
is, only in the last step of the search, we have to decompress the
rule, and process its components: $C \rightarrow 8~7$. The 8 is a
terminal symbol that lasts 1 time instant, and thus it is enough to
reach our target time instant. An 8 moves 1 column to the East and 1
to the North, which applied to the previous position (6,6) takes us
to the position (7,7).


The $MBR$ component aids during the computation of time-slice and
time-interval queries, as we will see soon. 

The additional elements enriching the rules are
compressed with an encoder designed for small integers (Directly
Addressable Codes, DAC) that support efficient access to any individual
value in the sequence \cite{BLN13}. To obtain better 
compression, the times of all the rules are compressed with one DAC, 
separately from the 3 pairs of coordinates of all the rules, which are 
compressed with another.

\vspace{-0.2cm}
\section{Querying}

\vspace{-0.15cm}
\subsubsection{Obtain the position of an object in a given time instant}

This query is solved by accessing the snapshot preceding the queried
time instant $t_q$, where we retrieve the position of the object at
the snapshot time instant. 
We then apply the movements of the log over this
position until we reach $t_q$. In the case of SCDC compression, we
follow the log decoding each codeword and applying  the relative
movement to the previous position. In the case of Re-Pair, we follow
the process described in the previous section.

\vspace{-0.35cm}
\subsubsection{Obtain the trajectory of an object between two time instants}

First, we obtain the position of the object in the start time
instant $t_s$, using the same algorithm of the previous query;
then we apply the movements of the log until reaching the end time
instant $t_e$.  In this case, when using GraCT, we have to
decompress $C$ to recover $I$, since only with $I$ we are capable of
describing the trajectory in detail, and thus we cannot take advantage
of the enriched nonterminal data. Therefore this query is
more time-consuming than the previous one for GraCT, and scdcCT takes
over.

\vspace{-0.35cm}
\subsubsection{Time slice query}
Given a time instant $t_q$ and a window rectangle $r$ of the space,
this query returns the objects that lie within $r$ at time $t_q$, and their positions.
We can distinguish two cases. First, if $t_q$ corresponds to a
snapshot, we only need to traverse the $k^2$-tree until the leaves,
inspecting those nodes that intersect $r$. When we reach the leaves,
we know their position and can retrieve from the permutation the
objects that are in this area.

The second case occurs when $t_q$ is between two snapshots $s_i$ and
$s_{i+1}$. In this case, we inspect in $s_i$ a region  $r'$, which
is an enlargement of the query region $r$. Region $r'$ is defined
using using the fastest object of the dataset as an upper bound. 
Thus, $r'$ is the rectangle containing all the points from where we can
reach the region $r$ at $t_q$ if moving at maximum speed along some direction.
Then, from $s_i$, we only track the objects that are within $r'$ in the
snapshot, therefore limiting the objects to follow and not wasting time with
objects that do not have chances to be in the answer.
We follow the movements of those objects from $s_i$
using the log, until reaching $t_q$. We further prune the tracking as we
process the log: a candidate object may follow a direction that takes it 
away from region $r$, so we recheck the condition after every movement and
discard an object as soon as it loses the chance of reaching $r$ at time $t_q$.

The tracking of objects is performed with the same algorithm explained
for obtaining the position of an object in a given time instant, but
in the case of GraCT, when a non terminal in the log corresponds to a
rule that brings the object we are following from an instant before
to an instant after $t_q$, instead of decompressing the nonterminal,
we intersect the MBR of the rule with $r$, and disregard the object if
the intersection is empty. Otherwise we decompress the nonterminal into
two and try again until reaching $t_q$ or discarding the object.

\begin{figure}[t]
\begin{center}

\includegraphics[scale=0.44]{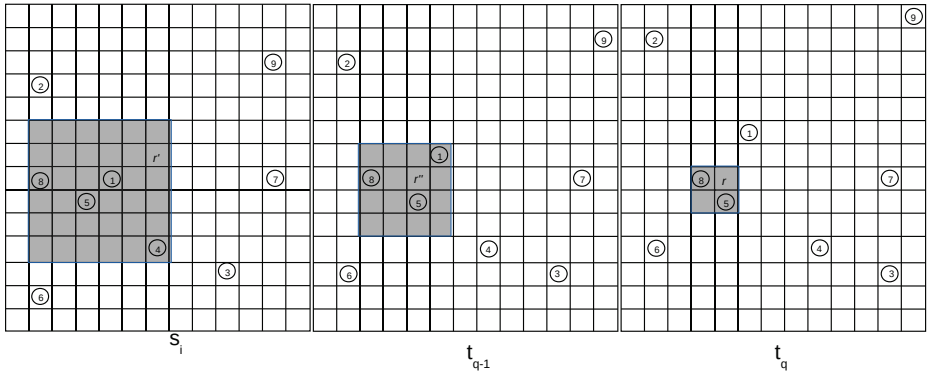}
\end{center}
\vspace*{-5mm}
\caption{Example of enlarged region $r'$ and query region $r$ in a time-slice query.}\label{timeSlice}
\end{figure}

Figure \ref{timeSlice} shows an example where we want to find the
objects that are located in $r$ at $t_q$. Assume that 
the fastest object can move only to an adjacent cell in
the period between two consecutive time instants. Let $s_i$ be the
last snapshot  preceding $t_q$ and let there be 2 time instants between 
$s_i$ and $t_q$. The left part of the figure shows the state of the objects 
at the time instant corresponding to $s_i$, in the middle to $s_i+1=t_q-1$, 
and in the right to $t_q$, where we show the region $r$. 
In $r'$ (shown on the left grid) we have four 
elements (1,4,5,8), which are candidates to be in $r$ at $t_q$, thus we
follow their log movements. In the middle grid, we show the
region $r''$ where the objects still have
chances to be within $r$ at $t_q$. Observe that, from the candidate
objects in $s_i$, object 4 has no further
chances to reach $r$, and thus it is not followed anymore. However,
object 1 still have chances, and therefore we keep tracking it. 

This query is affected by the  time elapsed between $s_i$ and $t_q$.
The farther away $s_i$ and $t_q$ are, the larger $r'$ will be, and
thus, we will have more candidate objects that have to be followed
through the log movements. In addition, with a large period between
$s_i$ and $t_q$, we have to traverse a longer portion of the log. To
alleviate this problem, if $t_q$ is closer to $s_{i+1}$ than to
$s_i$, we can start the search at $s_{i+1}$ and follow backwards the
movements of the log. For this backward traversal, we need to add
before each snapshot the last known position and its corresponding
time instant of the objects that are disappeared at the time instant
of the snapshot. This applies to both approaches scdcCT and GraCT.
Therefore, the maximum distance will be half of the distance between
two snapshots.

\vspace{-0.35cm}
\subsubsection{Time interval query}
In the time-slice query we have to know which objects are in $r$ at
the query time instant and their positions, but in the time-interval
query, the target is to know which objects were within $r$ at any
time instant of the  time interval $[t_s, t_e]$, specified in the
query. In this case, we  use the expanded region $r'$ again, which
is built as in the time-slice query, but using the time $t_e$.

Using SCDC, the objects within $r$ at $t_s$ are reported as part of
the solution, the other objects with chances at $t_s$ are followed
until they reach the region $r$, in which case they are added to the
answer; or when they move such that they lose the chance to
reach $r$ at $t_e$, in which case they are not followed anymore.

In GraCT, we process the log without decompressing nonterminal
symbols until the final time of a symbol in the log is equal or
larger than $t_s$. After this moment, for each symbol we read in the log,
until the object is selected or the next symbol in the log to read
starts after $t_e$, we follow the next procedure: For each log
symbol we check if the final point is inside $r$. If it is, the
object is selected. If not, and the MBR does not intersect $r$, we go on
to the next log symbol. If the final point is not in $r$ but the MBR
intersects $r$, we must apply the same procedure recursively to the pair of
symbols represented by the nonterminal, until the object is
selected or we process the whole nonterminal.

\section{Experimental Evaluation}

GraCT and scdcCT were coded in C++ and the experiments were run on a 1.70GHzx4 Intel Core i5 computer with 8GBytes of RAM and an operating system Linux of 64bits.

\vspace{-0.5cm}
\subsubsection{Datasets description and compression data}

We use a real dataset obtained from the site
\texttt{http://marinecadastre.gov/ais/}. The dataset provides the
location signals of 3,654 ships during a month. Every position
emitted by a ship is discretized into a matrix where the cell size
is $50 \times 50$ meters. With this data normalization, we obtain a matrix
with 100,138,325 cells, 36,775 in the $x$-axis and 2,723 in the $y$-axis.
Observe that our structure deals with object positions at regular
intervals, but in the dataset ship signals are emitted with
distinct frequencies, or they can send erroneous information.
Therefore, we preprocessed the signals to obtain regular time instants
every minute, thus discretizing the time into 44,642 minutes in
one month. With these settings, the original dataset occupies 501
MBs.

 \begin{table}[t]

 \begin{center}
\scriptsize
 \begin{tabular}{|l||r|r|r|r||r|r|r|r|}
  \hline
  {} & \multicolumn{4}{c||}{GraCT} & \multicolumn{4}{c|}{scdcCT} \\
  \hline
  Period & 120 & 240 & 360 & 720 & 120 & 240 & 360 & 720\\
  \hline
  \hline
   Size (MB) & 196.79 & 193.31 & 192.24 & 179.60 & 312.27 & 263.46 & 273.20& 282.95\\
  \hline
   Ratio & 39.27\% & 38.58\% & 38.36\% & 35.84\% & 62.32\% & 56.47\% & 54.52\% & 52.58\%\\
  \hline
  \hline
   Snapshot (MB) & 7.55 & 3.77 & 2.51 & 1.25 & 7.55 & 3.77 & 2.51 & 1.25\\
     & (3.83\%) & (1.95\%) & (1.31\%) & (0.70\%) & (2.42\%) & (1.33\%) & (0.92\%) & (0.48\%)\\
  \hline
  \hline
  Log (MB) & 189.25 & 189.54 & 189.73 & 178.34 & 304.73 & 279.18 & 270.69 & 262.20\\
   & (96.17\%) & (98.05\%) & (98.69\%) & (99.30\%) & (97.58\%) & (98.67\%) & (99.08\%) & (99.52\%) \\
  \hline

 \end{tabular}
\end{center}

 \caption{Compression ratio.}
 \label{table:experiment1}
\end{table}

 We built GraCT and scdcCT data structures over that
dataset using different snapshot distances, namely every 120, 240,
360, and 720 time instants. The construction time of the complete structure 
takes around 1 minute. Table \ref{table:experiment1} shows the
results of compression ratio\footnote{The size of the compressed
data structure as a percentage of the original dataset.}, where we
can see that GraCT obtains much better compression ratios than scdcCT. The rows
{\em snapshot} and {\em log} show the size of the snapshot and the
log as a percentage of the compressed data structure.  We can see that
the log is the most space demanding structure. As reference we compress the plain data with \textit{p7zip} and we obtain a compression ratio of 10,97\%, which is better than GraCT, however with this compressed data is impossible to answer any type of query.
 \subsubsection{Query types and answer times}

Table \ref{table:experiment2} shows the average answer times of 50
random queries of different types: {\em object} $t_q$ shows searches
for the position of an object in a given time instant, {\em trajectory} 
searches for
the trajectory followed by an object between two time instants, {\em
slice S} are time-slice queries that check for small regions ($367 \times 272$
cells) and {\em slice L} for large regions ($3677 \times 2723$ cells), 
{\em interval S} are time-interval queries with a small
region and a small time interval ($\frac{1}{10}$ of the snapshot
period) and {\em interval L} with large regions and a
large time interval ($\frac{1}{4}$ of the snapshot period).

\begin{table}[t]
\begin{center}
\scriptsize
\begin{tabular}{|l|r|r|r|r|r|r|r|r|}

 \hline
 {} & \multicolumn{4}{c|}{GraCT} & \multicolumn{4}{c|}{scdcCT} \\
 \hline
Period & 120 & 240 & 360 & 720 & 120 & 240 & 360 & 720\\
\hline
   Ratio & 39.27\% & 38.58\% & 38.36\% & 35.84\% & 62.32\% & 56.47\% & 54.52\% & 52.58\%\\
  \hline
 \hline
  Object $t_q$ & 0.0157 & \textbf{0.0169} & \textbf{0.0210} & \textbf{0.0229} & \textbf{0.0125} & 0.0169 & 0.0220 & 0.0246\\
 \hline
 Trajectory & 0.1582 & 0.1210 & 0.1130 & 0.1153 & \textbf{0.0881} & \textbf{0.0904} & \textbf{0.0900} & \textbf{0.0960}\\
 \hline
 Slice S & 1.5386 & \textbf{2.4241} & \textbf{2.9580} & \textbf{5.8788} & \textbf{1.3080} & 2.6712 & 4.0430 & 7.6636\\
 \hline
 Slice L & 1.7835 & \textbf{2.8074} & \textbf{3.6000} & \textbf{6.6430} & \textbf{1.5883} & 3.8615 & 5.1130 & 9.1384\\
 \hline
 Interval S & 2.4435 & \textbf{3.6005} & \textbf{5.0090} & \textbf{9.5635} & \textbf{1.4882} & 3.8612 & 5,8610 & 11.1765\\
 \hline
 Interval L & 2.7505 & \textbf{4.0847} & \textbf{6.1330} & \textbf{12.1832} & \textbf{1.7161} & 4.9578 & 9.8680 & 16.4471\\
 \hline
 \end{tabular}
\end{center}

 \caption{Time of different queries (ms). }
 \label{table:experiment2}
\end{table}

GraCT is the overall winner in all queries, except
in the {\em trajectory} query. This is expected, since to recover
the trajectory, GraCT has to decode all the symbols in $C$, given
that the enriched information in rules does not have the details of
the movements inside each rule. In the rest of the
queries the enriched information avoids in many cases to decode
the rules in $C$, and thus, since the logs in GraCT have far fewer
values than in scdcCT, the searches are faster. The exception is when
the size of the log between two snapshots is small, as the effect is
not noticeable. Notice that in this case the nonterminal
symbols in the grammar cannot represent arrays of more than 120
terminals.

\section{Conclusions}

We have presented a grammar based data structure for representing
 moving objects. It uses snapshots where
the objects are represented in the space using a $k^2$-tree and movement
logs that are grammar-compressed. The results of this
first experimental evaluation are very promising, as compression yields significant reductions in both space and time performance with respect to the baseline.

One reason why our space results are not even better is that the enriched
data pose a significant space overhead per nonterminal. We plan to improve
our representation by encoding these data in smarter ways. We also plan
to compare GraCT with state of the art indexes aimed at both time-slice
and time-interval queries, and  trajectories.

\bibliographystyle{splncs03}
\bibliography{bibliografia}

\end{document}